\documentclass[journal]{IEEEtran}

\usepackage[numbers, sort&compress]{natbib}

\usepackage{amsmath, amssymb, amsfonts}
\usepackage{empheq}
\usepackage{booktabs}
\usepackage{threeparttable}
\usepackage{multirow}
\usepackage{multicol}

\usepackage[thinlines, thiklines]{easybmat}

\usepackage[linesnumbered, ruled, vlined]{algorithm2e}

\usepackage{graphicx, subfigure, svg}  % illustration

\usepackage[english]{babel}

\usepackage{arydshln}

\usepackage{bbm}

\usepackage{enumerate}
\usepackage{paralist}
\usepackage{url}
\usepackage{color}

\usepackage{url}
\usepackage{hyperref}

\begin{document}

\renewcommand{\figurename}{Fig.}
\renewcommand{\tablename}{TABLE}

\title{Semi-Supervised Learning via Swapped Prediction \\ for Communication Signal Recognition}

% \author{Weidong~Wang, Hongshu~Liao, and Lu~Gan}

\author{
    \IEEEauthorblockN{Weidong~Wang, Hongshu~Liao, and Lu~Gan\vspace{-1em}}
    % \thanks{The work of Lu Gan was supported by Science and Technology on Electronic Information Control Laboratory. \it (Corresponding Author: Lu Gan).}
    \thanks{W. Wang, H. Liao, and L. Gan are with the School of Information and Communication Engineering, University of Electronic Science and Technology of China, Chengdu 611731, China. (e-mail: wwdong@std.uestc.edu.cn; hsliao@uestc.edu.cn; ganlu@uestc.edu.cn).}
}

\markboth{Communication Signal Recognition}
{Wang \MakeLowercase{\textit{et al.}}: Open-Set RF Fingerprinting}

\maketitle

\begin{abstract}
    Deep neural networks have been widely used in communication signal recognition and achieved remarkable performance, but this superiority typically depends on using massive examples for supervised learning, whereas training a deep neural network on small datasets with few labels generally falls into overfitting, resulting in degenerated performance. To this end, we develop a semi-supervised learning (SSL) method that effectively utilizes a large collection of more readily available unlabeled signal data to improve generalization. The proposed method relies largely on a novel implementation of consistency-based regularization, termed Swapped Prediction, which leverages strong data augmentation to perturb an unlabeled sample and then encourage its corresponding model prediction to be close to its original, optimized with a scaled cross-entropy loss with swapped symmetry. Extensive experiments indicate that our proposed method can achieve a promising result for deep SSL of communication signal recognition.
\end{abstract}


\IEEEpeerreviewmaketitle

\section{Introduction} \label{Section: Introduction}
\IEEEPARstart{D}{eep} learning has been widely used for communication signal recognition, including modulation recognition and radio frequency (RF) fingerprinting. Modulation recognition is an intermediate step between signal detection and demodulation to identify modulation types for received signals \cite{dobre2007survey}, which plays an important role in spectrum monitoring and interference identification. RF fingerprinting aims to distinguish different wireless transmitters by characterizing device-specific features (aka ``fingerprints") presented in their emitted signals \cite{soltanieh2020review}. As a promising non-password authentication technology, RF fingerprinting can greatly improve wireless security. Apart from modulation recognition and RF fingerprinting, any other target presented in communication signals can also be identified, which constitutes a generalized concept of communication signal recognition.

Conventional communication signal recognition largely relies on considerable domain knowledge and careful engineering to design suitable signal features, making various communication signal recognition tasks likely significant different. Since O'Shea \emph{et al.} \cite{o2016convolutional} successfully identified $11$ analog and digital modulation types using a simple convolutional neural network (CNN), a variety of methods based on deep learning have dominated this field \cite{mao2018deep}. The powerful non-linear representation of deep neural networks makes it possible to learn high-level features directly from raw signal data, which means that virtually all targets presented in communication signals can be effectively distinguished using such a general-purpose learning procedure. Hence, we believe that communication signal recognition has been unified to a certain extent.

Driven by an ideal number of high-quality examples, communication signal recognition based on deep learning has demonstrated its remarkable performance \cite{o2016convolutional, riyaz2018deep, zhang2022deep}. However, it is often difficult to obtain sufficient signal samples in practice. The biggest challenge comes from signal annotation, which requires much domain knowledge. The increasing number of signal captures can further complicate such annotation at great cost and time consumption. Training a deep neural network on small datasets often falls into overfitting and thus significantly degrades its generalization. One promising way to solve this issue is semi-supervised learning (SSL) \cite{zhu2022introduction}, which seeks to improve generalization by leveraging massive unlabeled data that are more easily available.

The recent studies on deep SSL are diverse, but those with consistency-based regularization have shown to work well in many fields \cite{oliver2018realistic}, including communication signal recognition. In a nutshell, consistency-based regularization encourages a model to give invariant predictions against any small perturbations applied to its input samples or hidden states. Different implementations of consistency-based regularization typically come with two key aspects. One is where and how to build perturbations. The other is how to measure such similarity between perturbed outputs and their original ones. The first issue has been well addressed in our previous work \cite{wang2023semi}, where we proposed a composite data augmentation scheme specifically designed for communication signals and leveraged it to perturb unlabeled signal samples, which can yield better results than noise injection.

The second aspect has been less investigated. Prior works \cite{laine2016temporal, tarvainen2017mean, miyato2018virtual,dong2021ssrcnn, fu2022novel, chen2022semi} conventionally consider popular distance measures like mean square error (MSE) or Kullback-Leibler (KL) divergence. In particular, some semi-supervised methods \cite{sohn2020fixmatch, fu2023semi, wang2023semi} utilize a pseudo-labeling procedure to sharpen predictions with relatively high confidence of unlabeled data into artificial labels. The similarity then can be directly calculated using a standard cross-entropy loss. This work analyzes different similarity measures and proposes a scaled cross-entropy loss with swapped symmetry. In conjunction with sample perturbation by strong data augmentation, it constitutes a novel implementation for consistency-based regularization, termed Swapped Prediction.

In sharp contrast to simply using MSE or KL divergence, ours not only ensures consistency but also produces more confident predictions due to entropy minimization. Meanwhile, it can effectively alleviate those incorrect but high-confidence model predictions from guiding wrong optimization directions, given that loss scaling is introduced. Along with Swapped Prediction, we introduce exponential moving averages (EMA) to improve performance and stability further, constituting an efficient semi-supervised algorithm for communication signal recognition named SS-CSR. The main contributions are summarized as follows:
\begin{itemize}
    \item  A novel implementation for consistency-based regularization termed Swapped Prediction is proposed.
    \item  An efficient semi-supervised algorithm for communication signal recognition is developed.
\end{itemize}

The proposed method for deep SSL of communication signal recognition is verified using a series of experiments on both simulated and real-world signal datasets. The experimental results demonstrate that our proposed method is far superior to other competing ones and only requires a small amount of labeled data to reach almost equivalent performance to full supervision.

\vspace{11pt}

The rest of this paper is organized as follows. The background and related work are introduced in Section~\ref{Section: Background and Related Work}. The proposed method is detailed in Section~\ref{Section: Methodology}. The experiments and related results are given in Section~\ref{Section: Experiments and Results} before concluding in Section~\ref{Section: Conclusion}.

\section{Background and Related Work} \label{Section: Background and Related Work}
Research on semi-supervised learning (SSL) has been ongoing for decades. There have been a variety of SSL methods, such as semi-supervised support vector machines (S3VM) \cite{bennett1998semi}, label propagation \cite{zhu2002learning}, and co-training \cite{blum1998combining}. See \cite{chapelle2006semi, zhu2022introduction} for a comprehensive overview of these conventional SSL methods. In recent years, deep neural networks have been demonstrated to achieve human- or beyond-human-level performance on certain supervised tasks (e.g., image classification) \cite{lecun2015deep}, leveraging a large collection of labeled data. To reduce such need for large-scale annotated datasets, it becomes increasingly attractive to train deep neural networks using a semi-supervised setting, commonly known as deep semi-supervised learning.

The most representative in deep SSL is a family of methods with consistency-based regularization \cite{sajjadi2016regularization, oliver2018realistic}, also known as consistency training, which yields a perturbed output through certain stochastic perturbations. The perturbed output is then enforced to be close to its original as a constraint. The related methods in this category are very diverse due to different ways of perturbations. In particular, some holistic solutions, such as FixMatch \cite{sohn2020fixmatch} and its variants, still rely on consistency-based regularization but simultaneously consider other techniques (e.g., pseudo-labeling) to improve generalization. There are many other methods in deep SSL. Some of them are based on generative models \cite{kingma2014semi, li2017triple, dong2019margingan}. The general idea is to exploit variational autoencoders (VAE) or generative adversarial networks (GAN) for learning sample distributions, which theoretically can produce more examples to help improve generalization. In recent years, self-supervised learning has flourished \cite{ericsson2022self}. Several methods based on contrastive representation learning lead to a promising result \cite{zhai2019s4l, chen2020big}. A comprehensive review of modern SSL methods can be found in \cite{chen2022semi, yang2022survey}.

This work investigates deep SSL for communication signal recognition. A brief review of relevant developments is conducted. Earlier, O'Shea \emph{et al.} \cite{o2017semi} employed a convolutional autoencoder (CAE) to learn low dimensional embedded representations from unlabeled signal data. The encoder part is then frozen and concatenated to a linear classifier, fine-tuned with a small number of labeled samples. However, such representations, learned by minimizing reconstruction errors, are often not strongly discriminative and thus not necessarily applicable to classification. Similarly, Liu \emph{et al.} \cite{liu2021self} also considered a two-stage solution but leveraged SimCLR \cite{chen2020simple}, a famous contrastive representation learning framework, to learn more discriminative representations, which can lead to a promising result. It is worth noting that contrastive representation learning relies heavily on data augmentation. Nevertheless, Liu \emph{et al.} only considered signal rotation, which somewhat restricts their performance.

There are also some related studies based on GANs, including SCGAN \cite{li2018generative}, E3SGAN \cite{zhou2020generative}, and BFE-CGAN \cite{tan2022semi}. However, unlike image data, it is challenging to synthesize realistic communication signal data when communication signals are high-order modulated or enjoy relatively small RF fingerprints \cite{wang2023radio}, making such methods perform poorly in practical cases. Recently, several methods adopted consistency-based regularization, including SSRCNN \cite{dong2021ssrcnn}, MAT \cite{fu2022novel}, DCR \cite{fu2023semi}, and CDA \cite{wang2023semi}, which can achieve highly competitive performance. These methods differ mostly in their respective implementations for consistency-based regularization, and later we will discuss them in detail and propose a novel implementation termed Swapped Prediction.

\section{Methodology} \label{Section: Methodology}
Formally, we are provided with a signal dataset $\mathcal{D} = \mathcal{S} \cup \mathcal{U}$ collected from $\mathrm{C}$ identifiable targets, where signal instances in $\mathcal{S}$ are labeled, i.e., $\mathcal{S} = \{(\boldsymbol{x}_i, \, y_i)\}_{i=1}^{\mathrm{M} \times \mathrm{C}}$, and those in $\mathcal{U}$ are not, i.e., $\mathcal{U} = \{\boldsymbol{x}_j\}_{j=1}^{\mathrm{N} \times \mathrm{C}}$, typically $\mathrm{M} \ll \mathrm{N}$. Note that we also refer to a signal instance with its label as a signal example. The core problem is how to utilize $\mathcal{U}$ to help a given deep model $f_\theta$ learning on $\mathcal{S}$ for communication signal recognition.

\subsection{Preliminaries}
Almost all modern semi-supervised algorithms employ a combined loss like
\begin{equation}
    \mathcal{L} = \mathcal{L}_\mathrm{s} + \lambda \mathcal{L}_\mathrm{u}
\end{equation}
where $\mathcal{L}_\mathrm{s}$ denotes a supervised objective, $\mathcal{L}_\mathrm{u}$ represents an unsupervised component, and $\lambda$ is a penalty factor that balances these two terms. This unsupervised term is often considered a regularization form that enables leveraging unlabeled data to improve generalization. As a matter of fact, many semi-supervised algorithms only differ in their respective regularization terms. In particular, consistency-based regularization \cite{sajjadi2016regularization, oliver2018realistic} is well acclaimed due to its SOTA performance.

Consistency-based regularization is in line with cluster assumption. Specifically, a sample will not easily change its belonging class after being slightly perturbed because those data points with different labels are typically separated by low-density regions. The trained model should have close predictions for an unlabeled sample and its perturbed version. The sample perturbation is not limited to various stochastic perturbations that act directly on samples but can also be indirectly achieved by perturbing a model itself, aka ``model perturbation". It is not difficult to see that consistency-based regularization has two essentials: (1) how to perturb an unlabeled sample; (2) how to enforce consistency of predictions for an unlabeled sample against its perturbed version. Different implementations for consistency-based regularization also differ mostly in these two aspects.

For issue (1), one can add random noise or consider other data augmentations, even indirectly achieved by some stochastic model perturbations (e.g., dropout). In deep SSL for communication signal recognition, SSRCNN by Dong \emph{et al.} \cite{dong2021ssrcnn} added simple noise like Gaussian with zero means, and it can also be adversarial noise \cite{miyato2018virtual} that MAT \cite{fu2022novel} has adopted, often leading to better performance. As introduced earlier, issue (1) has been well addressed by our past work \cite{wang2023semi} that proposed a composite data augmentation scheme specially designed for communication signals, whose effectiveness has been demonstrated in both supervised and semi-supervised learning. Hence, we shall uniformly leverage this composite data augmentation scheme to perturb unlabeled signal samples and then focus on issue (2).

Given an unlabeled signal sample $\boldsymbol{x} \in \mathcal{U}$, we can obtain its two outputs, $f_{\theta}\left(\boldsymbol{x}\right)$ and $f_\theta\left(\operatorname{g}\left(\boldsymbol{x}\right)\right)$, where $\operatorname{g}\left(\cdot\right)$ represents a data augmentation operation. The difference between $f_{\theta}\left(\boldsymbol{x}\right)$ and $f_\theta\left(\operatorname{g}\left(\boldsymbol{x}\right)\right)$ is then minimized to enforce consistency. To our knowledge, many existing methods adopt MSE or KL divergence as such similarity measures, as follows.
\begin{equation}
    \lVert f_{\theta}(\boldsymbol{x}) - f_{\theta}(\operatorname{g}(\boldsymbol{x})) \rVert^2
\end{equation}
and
\begin{equation}
    D_{\scriptscriptstyle \mathrm{KL}}\left(f_{\theta}\left(\boldsymbol{x}\right) \, \Vert \, f_\theta\left(\operatorname{g}\left(\boldsymbol{x}\right)\right)\right)
\end{equation}
Note that KL divergence can also be replaced by its symmetric form, i.e., Jensen-Shannon (JS) divergence, which makes no substantial difference. The KL divergence between two given distributions $\boldsymbol{p}$ and $\boldsymbol{q}$ is defined as
\begin{equation}
    D_{\scriptscriptstyle \mathrm{KL}}\left(\boldsymbol{p} \, \Vert \, \boldsymbol{q}\right) = - \sum_k p_k \log \frac{q_k}{p_k}
\end{equation}

Inspired by FixMatch \cite{sohn2020fixmatch}, a SOTA semi-supervised algorithm for image classification, both our past work CDA \cite{wang2023semi} as well as recent work DCR by Fu \emph{et al.} \cite{fu2023semi} have incorporated pseudo-labeling so that they can calculate such similarity between $f_{\theta}\left(\boldsymbol{x}\right)$ and $f_\theta\left(\operatorname{g}\left(\boldsymbol{x}\right)\right)$ directly with cross-entropy, i.e.,
\begin{equation}
    H\left(\tilde{y}, \, f_{\theta}\left(\operatorname{g}\left(\boldsymbol{x}\right)\right)\right) \cdot \mathbbm{1}_{\max\left(f_{\theta}\left(\boldsymbol{x}\right)\right) \ge \tau}
\end{equation}
where $\tau$ is a probability threshold used to only retain those predictions with high confidence, and $\tilde{y}$ represents pseudo labels sharpened from $f_{\theta}\left(\boldsymbol{x}\right)$, given by
\begin{equation}
    \tilde{y} = \mathop{\arg\mathop{\max}\limits_{c}}\left(f_{\theta}\left(\boldsymbol{x}\right)\right) \quad \text{if} \ \max\left(f_{\theta}\left(\boldsymbol{x}\right)\right) \ge \tau
\end{equation}
Note that $y \in \left\{1, \, 2, \, \ldots, \, \mathrm{C}\right\}$ or $\tilde{y}$ used here directly represents their corresponding one-hot encodings. The cross-entropy between two given distributions $\boldsymbol{p}$ and $\boldsymbol{q}$ is defined as
\begin{equation}
    H\left(\boldsymbol{p}, \, \boldsymbol{q}\right) = - \sum_k p_k \log q_k
\end{equation}

\subsection{Swapped Prediction}
Although there have been so many implementations for consistency-based regularization, they all have certain shortcomings. As previously stated, we focus on how to calculate such similarity between $f_{\theta}\left(\boldsymbol{x}\right)$ and $f_\theta\left(\operatorname{g}\left(\boldsymbol{x}\right)\right)$. An intuitive but not necessarily comprehensive analysis is given as follows. First, we do not recommend using MSE because its loss changes with output probabilities are relatively small, leading to a much smaller penalty for inconsistent cases than log-like losses. The penalty factor for loss balancing (i.e., $\lambda$) is also difficult to adjust, whereas using KL divergence or cross-entropy usually does not require such a penalty factor, or said $\lambda = 1.0$ is appropriate. According to a lot of practical experience, we also do not recommend incorporating pseudo-labeling. The unlabeled samples with incorrect pseudo labels can lead to a wrong optimization direction. Furthermore, only a small portion of unlabeled samples are annotated through pseudo-labeling during training, even setting a relatively low probability threshold. The information utilization of pseudo-labeling is relatively insufficient. The remaining is whether to choose KL divergence or use cross-entropy directly.

It is not difficult to deduce that
\begin{equation}
    H\left(\boldsymbol{p}, \, \boldsymbol{q}\right) = H\left(\boldsymbol{p}\right) + D_{\scriptscriptstyle \mathrm{KL}}\left(\boldsymbol{p} \, \Vert \, \boldsymbol{q}\right)
\end{equation}
Substitute into supervised classification, we have:
\begin{equation}
    H\left(y, \, f_\theta\left(\operatorname{g}\left(\boldsymbol{x}\right)\right)\right) = H\left(y\right) + D_{\scriptscriptstyle \mathrm{KL}}\left(y \, \Vert \, f_\theta\left(\operatorname{g}\left(\boldsymbol{x}\right)\right)\right)
\end{equation}
Since $H\left(y\right)$ is known, it is equivalent to a constant. In this case, there is no substantial difference between minimizing cross-entropy and minimizing KL divergence. In practice, one often prefers to adopt cross-entropy because it is easier to calculate. However, when it comes to consistency-based regularization, these two optimization objectives are no longer equivalent because $y$ has become $f_{\theta}\left(\boldsymbol{x}\right)$, which is unknown and needs to be optimized. If optimizing with KL divergence, one can only ensure that $f_{\theta}\left(\boldsymbol{x}\right)$ and $f_{\theta}\left(\operatorname{g}\left(\boldsymbol{x}\right)\right)$ tend to be consistent, while optimizing with cross-entropy can make $f_{\theta}\left(\boldsymbol{x}\right)$ further sharpened since it minimizes  $H\left(f_{\theta}\left(\boldsymbol{x}\right)\right)$, i.e., entropy minimization. Such high-confidence predictions are exactly what a classification task expects. Hence, we argue that optimizing with cross-entropy should be slightly better than KL divergence in theory.

\begin{figure}[htb]
    \centering
    \includegraphics[scale=0.625]{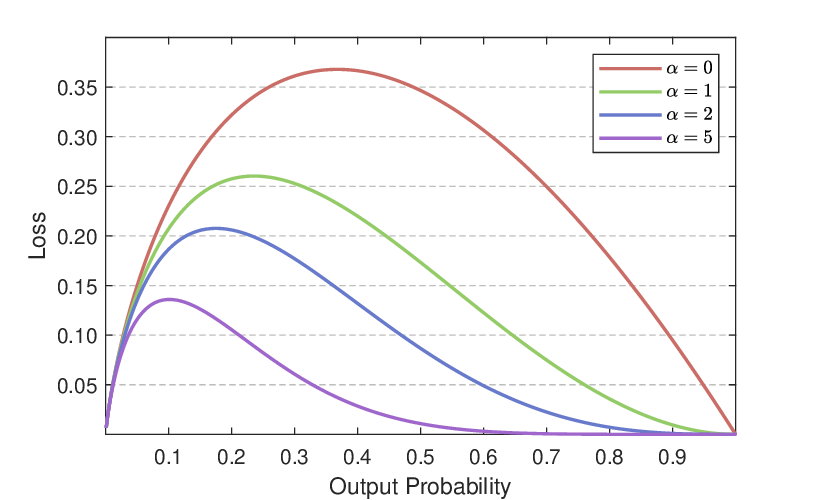}
    \caption{The scaled cross-entropy loss is plotted under different $\alpha$ settings, given that $f_{\theta}\left(\boldsymbol{x}\right) \approx f_{\theta}\left(\operatorname{g}\left(\boldsymbol{x}\right)\right)$. The output probability $p$ for each class by random guess is $1/\mathrm{C}$, so we only need to focus on $p \ge 1/\mathrm{C}$.}
    \label{Figure: Scaled Cross-Entropy}
\end{figure}

However, we also notice that many unlabeled samples could be wrongly predicted in practice. The samples that are wrongly given high confidence can result in a wrong optimization direction, just like pseudo-labeling. This instead deteriorates generalization performance. To address this issue, inspired by vanilla Focal Loss \cite{lin2017focal}, we introduce a scaling factor $\alpha$ for standard cross-entropy, defined by
\begin{equation}
    H_{\alpha} \left(\boldsymbol{p}, \, \boldsymbol{q}\right) = - \sum_k (1-p_k)^{\alpha} p_k \log q_k
\end{equation}
where $\alpha \ge 0$. The standard cross-entropy is taken by $\alpha = 0$. This loss scaling significantly reduces loss contributions from high-confidence predictions, as shown in Fig. \ref{Figure: Scaled Cross-Entropy}, and thus makes our training procedure focus on those low-confidence predictions. Adjusting $\alpha$ allows one to adapt to various signal datasets with different sample conditions flexibly.

Moreover, it should be important to point out that $f_{\theta}\left(\boldsymbol{x}\right)$ and $f_{\theta}\left(\operatorname{g}\left(\boldsymbol{x}\right)\right)$ are actually equivalent in their respective roles. This is different from supervised classification in which $y$ and $f_{\theta}\left(\boldsymbol{x}\right)$ have their unambiguous positions when calculating a cross-entropy loss derived from maximum likelihood. To make $f_{\theta}\left(\boldsymbol{x}\right)$ and $f_{\theta}\left(\operatorname{g}\left(\boldsymbol{x}\right)\right)$ approximate each other, instead of $f_{\theta}\left(\operatorname{g}\left(\boldsymbol{x}\right)\right)$ unidirectionally approaching $f_{\theta}\left(\boldsymbol{x}\right)$, we need to consider a symmetric form:
\begin{equation}
    \frac{H_{\alpha}\left(f_{\theta}\left(\boldsymbol{x}\right)\right), \, f_{\theta}\left(\operatorname{g}\left(\boldsymbol{x}\right)\right) + H_{\alpha}\left(f_{\theta}\left(\operatorname{g}\left(\boldsymbol{x}\right)\right), \, f_{\theta}\left(\boldsymbol{x}\right)\right)}{2}
\end{equation}
which can lead to certain improvements in performance and stability, given that using an asymmetrical single-side form sometimes may not converge effectively or easily fall into bad minima. For clarity, we refer to this novel implementation for consistency-based regularization as ``Swapped Prediction".

% \begin{equation}
%     \frac{1}{2} \{ H_{\alpha}\left(f_{\theta}\left(\boldsymbol{x}\right)\right), \, f_{\theta}\left(\operatorname{g}\left(\boldsymbol{x}\right)\right) + H_{\alpha}\left(f_{\theta}\left(\operatorname{g}\left(\boldsymbol{x}\right)\right), \, f_{\theta}\left(\boldsymbol{x}\right)\right) \}
% \end{equation}

% \begin{equation}
%     \begin{split}
%         \frac{1}{2} \{
%         & H_{\mathrm{scaled}}\left(f_{\theta}\left(\boldsymbol{x}\right)\right), \, f_{\theta}\left(\operatorname{g}\left(\boldsymbol{x}\right)\right) + \\ & H_{\mathrm{scaled}}\left(f_{\theta}\left(\operatorname{g}\left(\boldsymbol{x}\right)\right), \, f_{\theta}\left(\boldsymbol{x}\right)\right)
%         \}
%     \end{split}
% \end{equation}

% \begin{equation}
%     \ell_\mathrm{s}\left(f_{\theta}\left(\operatorname{g}\left(\boldsymbol{x}\right)\right), \, y\right) = H\left(y, \, f_{\theta}\left(\operatorname{g}\left(\boldsymbol{x}\right)\right)\right)
% \end{equation}
% and
% \begin{equation}
%     \begin{split}
%         \ell_\mathrm{u}\left(f_{\theta}\left(\operatorname{g}\left(\boldsymbol{x}\right)\right), \, f_{\theta}\left(\boldsymbol{x}\right)\right) = \frac{1}{2} \{
%         H_{\mathrm{scaled}}\left(f_{\theta}\left(\boldsymbol{x}\right)\right), \, f_{\theta}\left(\operatorname{g}\left(\boldsymbol{x}\right)\right) \\ + H_{\mathrm{scaled}}\left(f_{\theta}\left(\operatorname{g}\left(\boldsymbol{x}\right)\right), \, f_{\theta}\left(\boldsymbol{x}\right)\right)
%         \}
%     \end{split}
% \end{equation}

\subsection{Strong Data Augmentation}
\begin{figure*}[htb]
    \centering
    \includegraphics[scale=0.68]{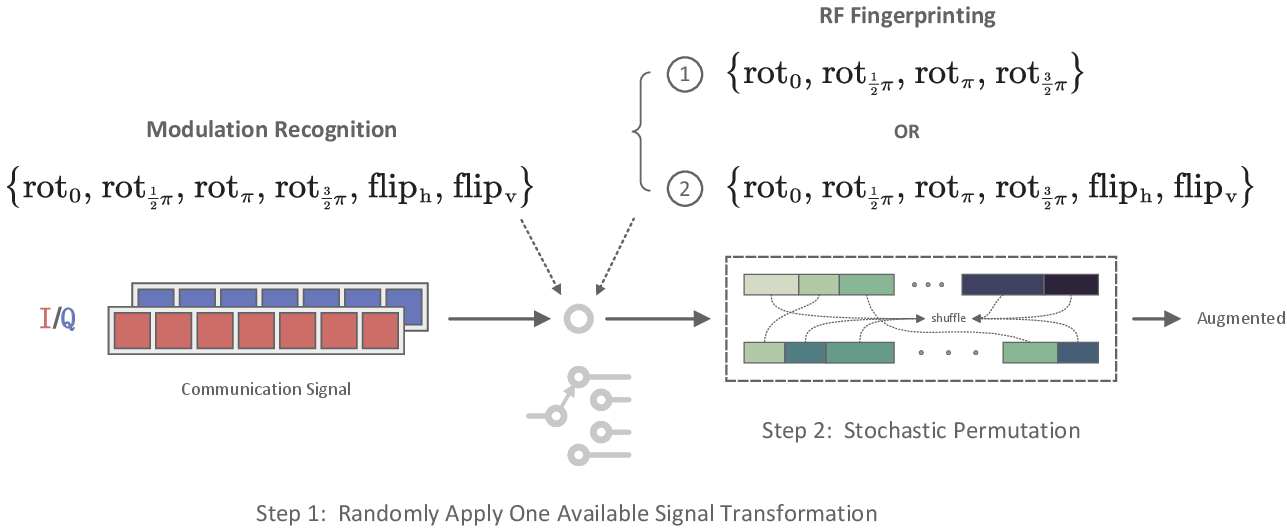}
    \caption{A composite data augmentation scheme specially designed for communication signals.}
    \label{Figure: Composite Data Augmentation}
\end{figure*}

Data augmentation is very important for deep SSL but is often domain-specific. As stated, our past work \cite{wang2023semi} has proposed an effective data augmentation scheme for communication signals. As illustrated in Fig. \ref{Figure: Composite Data Augmentation}, it is a two-step composite operation that randomly selects one from a set of available signal transformations to apply and then performs stochastic permutation. More specifically, we have two types of signal transformations: rotation and flipping \cite{huang2019data}. Given a complex-valued signal $\boldsymbol{s}$, its rotated version by angle $\vartheta$ is:
\begin{equation}
    \operatorname{rot}_\vartheta: \quad \boldsymbol{s}_{\vartheta} = \boldsymbol{s} e^{j\vartheta}
\end{equation}
The signal flipping consists of horizontal flipping and vertical flipping, which are defined as follows:
\begin{equation}
    \displaystyle \left\{
    \begin{aligned}
        \operatorname{flip}_{\mathrm{h}}: \quad \boldsymbol{s}_\mathrm{h} & = \operatorname{conj}(-\boldsymbol{s}) \quad \text{// Flip Horizontally} \\
        \operatorname{flip}_{\mathrm{v}}: \quad \boldsymbol{s}_\mathrm{v} & = \operatorname{conj}(\boldsymbol{s}) \quad \text{// Flip Vertically}    \\
    \end{aligned} \right.
\end{equation}
where $\operatorname{conj}(\cdot)$ means taking conjugation. In modulation recognition, we can rotate a signal sample by $0^\circ$, $90^\circ$, $180^\circ$, $270^\circ$, i.e., $\left\{\operatorname{rot}_{0}, \, \operatorname{rot}_{\frac{1}{2}\pi}, \, \operatorname{rot}_{\pi}, \, \operatorname{rot}_{\frac{3}{2}\pi}\right\}$. Here plus two flipping operations, we can use a total of $6$ signal transformations.

The situation becomes relatively complicated in RF fingerprinting. In general, given a group of very similar devices with completely identical signal parameters configured, we can only use signal rotation, as flipping theoretically can destroy RF fingerprints, introducing additional noise. The angle for rotation can be customized individually depending on what modulation type is used by each device. Nevertheless, in most cases, we still choose $\left\{\operatorname{rot}_{0}, \, \operatorname{rot}_{\frac{1}{2}\pi}, \, \operatorname{rot}_{\pi}, \, \operatorname{rot}_{\frac{3}{2}\pi}\right\}$ as a more adaptable solution. Moreover, in many practical cases, except for RF fingerprints, it is also possible to identify different devices with other valid information, including their associated signal parameters. For example, different models of devices may adopt different wireless protocols and use different modulation types, which may sometimes serve as identification criteria to distinguish different individuals, even though these characteristics are not strictly device-specific. In this case, it is still necessary to consider flipping. Hence, we recommend only using rotation or a combination of rotation and flipping as data augmentation in RF fingerprinting, depending on what signal dataset we use.

The so-called stochastic permutation shall split a signal sample into multiple segments, then shuffle and re-concatenate them into a new sequence. For example, split into $k$ segments, often termed ``$k$-segmented stochastic permutation". In general, we should consider a bigger $k$ in modulation recognition while a smaller $k$ in RF fingerprinting, or directly taken by default, $k = 2$. See our past work \cite{wang2023semi} for more details.

Then, we analyze why this composite data augmentation scheme can play very effective role in SSL for communication signal recognition. The sample distribution can actually be described using a finite set of support points. For simplicity, our discussion is based on a single class. Suppose that the sample distribution of the class in the sample space can be described by $n$ support points, i.e., $\Omega = \{\omega_1, \, \omega_2, \, \dots, \, \omega_n\}$, and $\operatorname{distance}\left(\cdot\right)$ is a valid distance definition in this sample space. Given a sample $\boldsymbol{x}$ that belongs to the class and the corresponding support point $\omega \in \Omega$, if $\operatorname{distance}\left(\boldsymbol{x}, \, \omega\right) \le \epsilon$, then $\boldsymbol{x}$ is said to belong to the support point $\omega$, where $\epsilon$ is a smaller error coefficient. From the view of the sample space, $\boldsymbol{x}$ is a sample within a smaller range centered on $\omega$.

On this basis, the intensity of data augmentation can be defined, as follows. Formally, given a data augmentation operation $\mathcal{A}\left(\cdot\right)$, we can have a augmented version $\tilde{\boldsymbol{x}} = \mathcal{A}\left(\boldsymbol{x}\right)$. Suppose that $\tilde{\boldsymbol{x}}$ belongs to $\tilde{\omega}$, if $\tilde{\omega} = \omega$, or
\begin{equation}
    \operatorname{distance}\left(\boldsymbol{\tilde{x}}, \, \omega\right) \le \epsilon
\end{equation}
Then, we refer to $\mathcal{A}\left(\cdot\right)$ as ``weak augmentation", that is, $\mathcal{A}\left(\cdot\right)$ makes a very small change to samples, so that $\tilde{\boldsymbol{x}}$ and $\boldsymbol{x}$ are very close in their sample space, and both belong to the same support point; otherwise, $\mathcal{A}\left(\cdot\right)$ is called ``strong augmentation". For strong augmentation, there are two cases:
\begin{itemize}
    \item $\tilde{\omega} \in \Omega$, that is, although the augmented version no longer shares the same support point with the original sample, it still belongs to the same class.
    \item $\tilde{\omega} \notin \Omega$, that is, the augmented version no longer belongs to the class, which means that out-of-distribution samples have been generated.
\end{itemize}
In most cases, out-of-distribution samples should not be used for classification, which brings the confirm bias and thus mislead inference. Obviously, noise injection is typically weak augmentation. Although as long as such injected noise is large enough, it can greatly change a sample, but this usually produces out-of-distribution samples. Too much noise will drown out the original features and cannot produce positive gains. In contrast, including rotation, flipping, stochastic permutation, and our composite data augmentation scheme, all belong to strong augmentation. More importantly, our composite data augmentation scheme hardly changes the sample distribution. This definition about ``weak" and ``strong" differs from \cite{sohn2020fixmatch}. The latter only depends on such relative complexity between different data augmentation operations.

\begin{figure}[htb]
    \centering
    \subfigure[before augmentation]
    {
        \centering
        \includegraphics[width=0.35\textwidth]{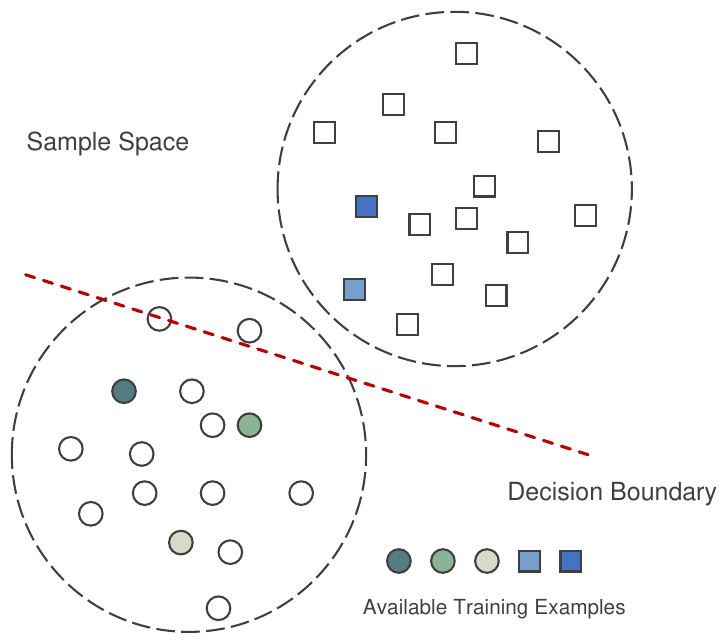}
    }
    \\
    \subfigure[after augmentation]
    {
        \centering
        \includegraphics[width=0.35\textwidth]{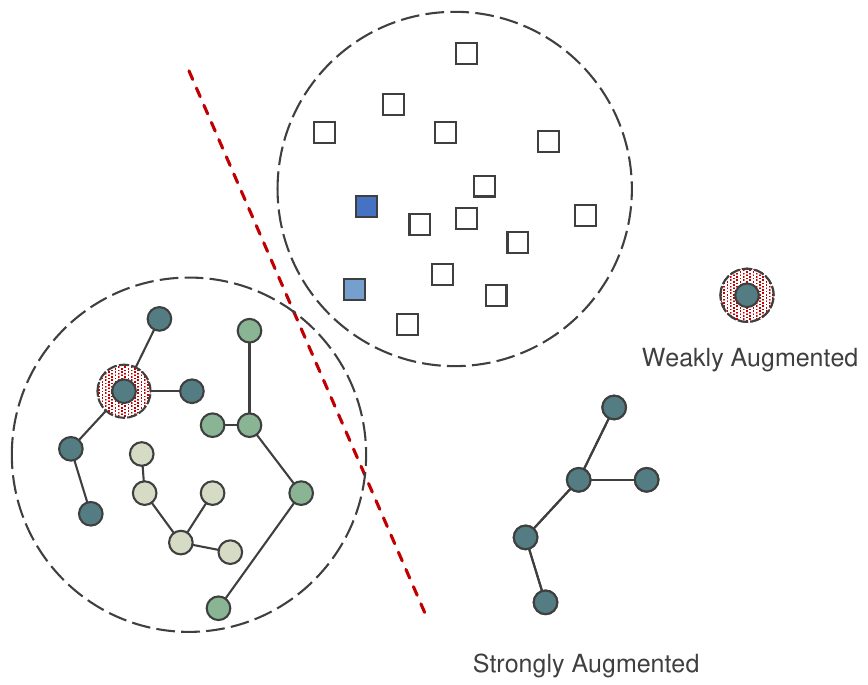}
    }
    \caption{Illustration of Sample Space before and after data augmentation.}
    \label{Figure: Sample Space}
\end{figure}

As shown in Fig. \ref{Figure: Sample Space}, when training with a very limited number of examples, the scope of the sample space that the model can perceive is restricted, or only part of the support points are seen by the model, which makes the model unable to find a reliable decision boundary. This is an intuitive explanation to overfitting. The proposed composite data augmentation belongs to strong augmentation, and almost does not change the original sample distribution, which can make the sample transform to another support point of the same class after data augmentation. Theoretically, as long as $\mathcal{A}\left(\cdot\right)$ produces enough changes, it can cover more support points, which means that it is possible to make the model perceive other support points through the proposed composite data augmentation, and essentially expand the range of sample space that the model can perceive. In contrast, weak data augmentation such as noise injection can only make the sample float in a small range around itself, and can only play a role in smoothing the decision boundary, but cannot substantially expand the range of sample space that the model can perceive through, resulting in a very limited corresponding generalization improvement. Through the above analysis, it is not difficult to know that our composite data augmentation scheme can fundamentally improve model generalization caused by insufficient training examples.

This characteristic also benefits consistency-based regularization. When the model makes consistent prediction to $\boldsymbol{x}$ and $\tilde{\boldsymbol{x}}$, which is essentially to make $\omega$ and $\tilde{\omega}$ close to each other in the learned embedding feature space, that is, to produce aggregation. If $\mathcal{A}\left(\cdot\right)$ has randomness, it is not difficult to know that the more changes $\mathcal{A}\left(\cdot\right)$ produces, the more $\tilde{\omega}$ covers the remaining support points except for  $\omega$, the more significant the aggregation effect finally presented in the learned feature space. Theoretically, if the randomness of $\mathcal{A}\left(\cdot\right)$ can make $\tilde{\omega}$ cover all the remaining support points, through consistent prediction, all samples of this class will be gathered together, which is equivalent to giving all unlabeled samples an implicit label. According to this analysis, when leveraging our composite data augmentation scheme to perturbed unlabeled signal samples, $\operatorname{rot}_{0}$ is recommended to discard.

\subsection{Exponential Moving Average}
To further improve generalization, we consider it relatively useful to maintain a moving average of all trainable model parameters during training. The moving average of $\theta$ is updated using an exponential decay $\gamma$ after each training step:
\begin{equation}
    \theta_{\scriptscriptstyle \mathrm{EMA}} \leftarrow \gamma \theta_{\scriptscriptstyle \mathrm{EMA}} + (1-\gamma) \theta
\end{equation}
The reasonable decay is close to $1.0$, typically in a multiple-nines range, e.g., $0.9$ and $0.99$, which means that we can determine this hyper-parameter at a very small cost.

An additional difference between conventional EMA and ours is that we apply training to $\theta_{\scriptscriptstyle \mathrm{EMA}}$, which replaces $\theta$ with $\theta_{\scriptscriptstyle \mathrm{EMA}}$ after each training epoch as a new optimization basis to obtain better stability. While conventional EMA often treats it as a constant regarding optimization that only provides a better trained model for final evaluation but actually does not impact training. This idea is essentially equivalent to an online model ensemble.

Although many other commonly used generalization improvements could also be integrated into deep SSL to improve performance further, most are not worth using in practice due to their additional hyper-parameters that are usually not easy to configure, not significant performance gains, or other possible restrictions. For example, we know that noise injection can play a certain regularization role. The size of such injected noise, however, is often difficult to configure precisely. It might vary with different data conditions, e.g., SNR and communication signal type. Unlike image data, communication signals are more noise-sensitive, especially for RF fingerprinting, where such minor signal distortions caused by RF fingerprints could easily be overwhelmed by noise. In contrast, we believe that EMA are so simple and applicable.

\subsection{Training in Semi-Supervised Fashion}
For convenience, we define
\begin{equation}
    \displaystyle \left\{
    \begin{aligned}
         & \ell_\mathrm{s}\left(\boldsymbol{p}, \, \boldsymbol{q}\right)              = \displaystyle H\left(\boldsymbol{q}, \, \boldsymbol{p}\right)                                                                                                \\
         & \ell_{\mathrm{u}}\left(\boldsymbol{p}, \, \boldsymbol{q}; \, \alpha\right) = \displaystyle \frac{1}{2} \left\{H_{\alpha}\left(\boldsymbol{q}, \, \boldsymbol{p}\right) + H_{\alpha}\left(\boldsymbol{p}, \, \boldsymbol{q}\right)\right\} \\
    \end{aligned} \right.
\end{equation}
In a nutshell, we leverage strongly augmented data for supervised training, i.e.,
\begin{equation}
    \mathcal{L}_\mathrm{s} = \displaystyle \frac{1}{\lvert\mathcal{S}\rvert} \sum_{\left(\boldsymbol{x}, \, y\right) \, \in \, \mathcal{S}} {\ell_\mathrm{s}\left(f_{\theta}\left(\operatorname{g}\left(\boldsymbol{x}\right)\right), \, y\right)}
\end{equation}
and simultaneously achieve consistency-based regularization by swapped prediction, i.e.,
\begin{equation}
    \mathcal{L}_\mathrm{u} = \displaystyle \frac{1}{\lvert\mathcal{U}\rvert} \sum_{\boldsymbol{x} \, \in \, \mathcal{U}} {\ell_\mathrm{u}\left(f_{\theta}\left(\operatorname{g}\left(\boldsymbol{x}\right)\right), \, f_{\theta}\left(\boldsymbol{x}\right); \, \alpha\right)}
\end{equation}
The overall training pipeline is summarized in Algorithm \ref{Algorithm: SSCSR}. For clarity, we name it semi-supervised communication signal recognition (SSCSR).

\begin{algorithm}
    \caption{Semi-Supervised Learning for Communication Signal Recognition}
    \label{Algorithm: SSCSR}
    \SetAlgoLined
    \For{{\rm each training epoch}}{
    \Repeat{{\rm Traverse $\mathcal{D}$}}{
    Sample $\mathcal{B}_{\mathrm{s}}$ and $\mathcal{B}_{\mathrm{u}}$ from $\mathcal{S}$ and $\mathcal{U}$, respectively. \\
    Forward propagation, calculate loss:\\
    $\mathcal{L}_{\mathrm{s}} = \displaystyle \frac{1}{\lvert\mathcal{B}_{\mathrm{s}}\rvert} \sum_{\left(\boldsymbol{x}, \, y\right) \, \in \, \mathcal{B}_{\mathrm{s}}} {\ell_{\mathrm{s}}\left(f_{\theta}\left(\operatorname{g}\left(\boldsymbol{x}\right)\right), \, y\right)}$ \\
    $\mathcal{L}_{\mathrm{u}} = \displaystyle \frac{1}{\lvert\mathcal{B}_{\mathrm{u}}\rvert} \sum_{\boldsymbol{x} \, \in \, \mathcal{B}_{\mathrm{u}}} {\ell_{\mathrm{u}}\left(f_{\theta}\left(\operatorname{g}\left(\boldsymbol{x}\right)\right), \, f_{\theta}\left(\boldsymbol{x}\right); \, \alpha\right)}$ \\
    \vskip 0.125cm
    $\mathcal{L} = \displaystyle \mathcal{L}_\mathrm{s} + \mathcal{L}_\mathrm{u}$ \\
    \vskip 0.125cm
    Back propagation, optimize model parameters based on SGD:\\
    $\theta \leftarrow \operatorname{SGD}\left(\nabla_{\theta}{\mathcal{L}}, \; \theta\right)$ \\
    \vskip 0.125cm
    Maintain EMA:\\
    $\theta_{\scriptscriptstyle \mathrm{EMA}} \leftarrow \gamma \theta_{\scriptscriptstyle \mathrm{EMA}} + \left(1-\gamma\right) \theta$ \\
    \vskip 0.125cm}
    Apply EMA:\\
    $\theta \leftarrow \theta_{\scriptscriptstyle \mathrm{EMA}}$ \\
    \vskip 0.125cm}
\end{algorithm}

% \begin{algorithm}
%     \label{Algorithm: SSCSR}
%     \caption{SS-CSR}
%     \SetAlgoLined
%     \For{{\rm each training epoch}}{
%     \vskip 0.125cm
%     \For{{\rm each training step}}
%     {
%     \vskip 0.125cm
%     Sample $\mathcal{B}_{\mathrm{s}}$ and $\mathcal{B}_{\mathrm{u}}$ from $\mathcal{S}$ and $\mathcal{U}$, respectively.
%     \vskip 0.125cm
%     $
%         \mathcal{L}_\mathrm{s} = \displaystyle \frac{1}{\lvert\mathcal{B}_{\mathrm{s}}\rvert} \sum_{\left(\boldsymbol{x}, \, y\right) \in \mathcal{B}_{\mathrm{s}}} {\ell_\mathrm{s}\left(f_{\theta}\left(\operatorname{g}\left(\boldsymbol{x}\right)\right), \, y\right)}
%     $ \\
%     \vskip 0.125cm
%     $
%         \mathcal{L}_\mathrm{u} = \displaystyle \frac{1}{\lvert\mathcal{B}_{\mathrm{u}}\rvert} \sum_{\boldsymbol{x} \in \mathcal{B}_{\mathrm{u}}} {\ell_\mathrm{u}\left(f_{\theta}\left(\operatorname{g}\left(\boldsymbol{x}\right)\right), \, f_{\theta}\left(\boldsymbol{x}\right)\right)}
%     $ \\
%     \vskip 0.125cm
%     $
%         \mathcal{L} = \mathcal{L}_\mathrm{s} + \mathcal{L}_\mathrm{u}
%     $ \\
%     \vskip 0.125cm
%     \textbf{Optimize}
%     $
%         \theta \leftarrow \text{SGD}(\nabla_{\theta}{\mathcal{L}}, \; \theta)
%     $ \\
%     \vskip 0.125cm
%     \textbf{Update EMA}
%     $
%         \theta_{\scriptscriptstyle \mathrm{EMA}} \leftarrow \gamma \theta_{\scriptscriptstyle \mathrm{EMA}} + (1-\gamma) \theta
%     $ \\
%     \vskip 0.125cm
%     }
%     \vskip 0.125cm
%     \textbf{Replace} $\theta \leftarrow \theta_{\scriptscriptstyle \mathrm{EMA}}$ \\
%     \vskip 0.125cm
%     }
% \end{algorithm}

\section{Experiments and Results} \label{Section: Experiments and Results}
In this section, a series of experiments are conducted to evaluate our proposed method comprehensively.

\subsection{Data Preparation}
The experiments adopt both simulated and real-world signal data. More specifically, all our ablation experiments will adopt communication signal simulation since it can facilitate precise control of various experimental conditions and exclude other possible interferences, while such comparisons with other competitive methods will consider two public signal datasets, RadioML 2018.01A \cite{o2018over} and WIDEFT \cite{siddik2021wideft}, for convincing results. The related signal datasets are given as follows.

\vspace{11pt}

\noindent\textbf{Simulation} The signal adopts QPSK modulation with pulse-shaping by a square-root raised cosine filter of roll-off factor $0.35$. The additive white Gaussian noise (AWGN) channel of $\mathrm{SNR} = 18 \text{ dB}$ is considered. The length of each sample is $1024$, with $8 \times$ oversampling. Note that we consider non-linear power amplification to yield specific RF fingerprints for RF fingerprinting. See our previous work \cite{wang2023semi} for more details. The signal data with a total of $10$ simulated devices is randomly divided into a training set, validation set, and test set in a proportion of $3:1:1$, with $10000$ samples per class generated.

\vspace{11pt}

\noindent\textbf{RadioML 2018.01A} This is an open-source signal dataset, available on \href{https://www.deepsig.ai}{DeepSig}, and it contains $24$ digital and analog modulation types, including OOK, 4ASK, 8ASK, BPSK, QPSK, 8PSK, 16PSK, 32PSK, 16APSK, 32APSK, 64APSK, 128APSK, 16QAM, 32QAM, 64QAM, 128QAM, 256QAM, AM-SSB-WC, AM-SSB-SC, AM-DSB-WC, AM-DSB-SC, FM, GMSK, OQPSK, each of which involves different SNRs, varying from $-20$ to $30$ dB with an interval of $2$ dB. There are $4096$ signal samples of length $1024$ under each SNR for every modulation type, where we only consider $\mathrm{SNR} = 10 \text{ dB}$ and randomly select $2096$ samples as a training set. The remaining for each class are equally divided into two groups, used for validation and evaluation, respectively.

\vspace{11pt}

\noindent\textbf{WIDEFT} This RF dataset is collected from $138$ real-world devices (e.g., smartphones, headsets, routers), available on \href{https://zenodo.org/record/4116383}{Zenodo}. The signal captures of all Apple Inc devices equipped with $2.4$ GHz WiFi are selected, with a total of $18$ devices. Each capture consists of $100$ bursts, i.e., $100$ bursts per device, subsequently divided into a training set, validation set, and test set in a proportion of $3:1:1$. Note that each signal burst is complete that consists of ON transient, steady-state portion, and OFF transient, and includes $5000$ sampling points before and after. The steady-state portion is long enough to be sliced into multiple samples, and we randomly slice each burst into $50$ signal samples of length $1024$.

\vspace{11pt}

The training data is then further assigned as ``labeled" and ``unlabeled" data according to specific data conditions. The data condition like ``$\mathrm{A} + \mathrm{B}$" means $\mathrm{M} = \mathrm{A}$ and $\mathrm{N} = \mathrm{B}$.

\subsection{Implementation and Training Details}
\begin{figure}[htb]
    \centering
    \includegraphics[scale=0.82]{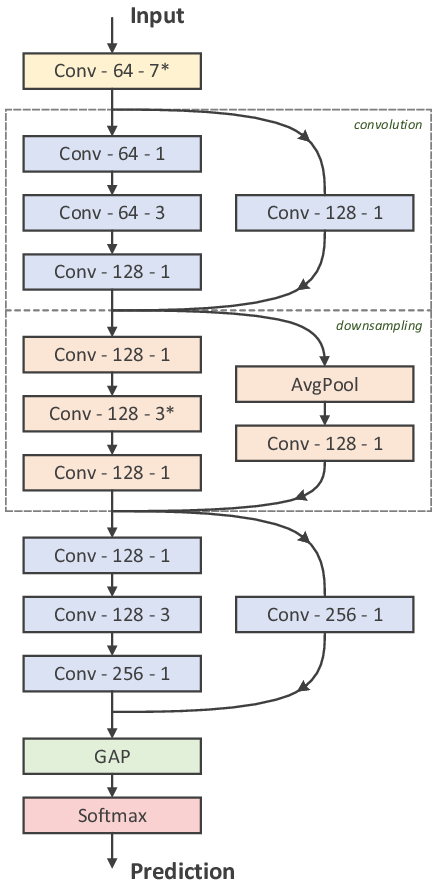}
    \caption{ResNet for communication signal recognition. The layer label ``Conv-$k$-$s$'' indicates that this convolutional layer has $k$ kernels of size $s$. The convolutions of size $3$ are all separable. The asterisk (*) indicates that this layer enjoys a stride of size $2$. Batch normalization (BN) \cite{ioffe2015batch} has been performed after each convolution, followed by activation using rectified linear units (ReLU) \cite{nair2010rectified}.}
    \label{Figure: Network}
\end{figure}

This work adopts a deep residual network (ResNet) \cite{he2016deep} to identify communication signals. As illustrated in Fig. \ref{Figure: Network}, it starts with a convolution layer, followed by a series of alternately stacked convolution and downsampling blocks, and ends with a classification layer. The initial convolution layer is used as an input stem and has $64$ kernels of size $7$ with a stride of $2$. The convolution and downsampling blocks are both implemented as residual blocks. The last classification layer performs global average pooling (GAP) and then yields a prediction using a dense layer activated by softmax.

The model is built with TensorFlow \cite{tensorflow2015} and then trained on a single NVIDIA RTX 4090 GPU utilizing an Adam \cite{kingma2014adam} optimizer for $380$ epochs. The batch size could affect training stability and convergence, and we configure it to $32\text{/}128$ in most cases, meaning that each training step will leverage $32$ labeled samples and $128$ unlabeled samples, a relatively optimal setting obtained through extensive trials. The initial learning rate is set to $0.001$. For each experimental setting (e.g. different data conditions), we show its best performance over $10$ consecutive trials.

\begin{table*}[htb]
    \renewcommand\arraystretch{1.15}
    \centering
    \caption{Comparison of Different Consistency Forms}
    \label{Table: Comparison of Different Consistency Forms}
    \resizebox{0.83\textwidth}{!}{
        \begin{threeparttable}
            \begin{tabular}{cccccc}
                \toprule
                \textbf{Consistency}
                 & $\boldsymbol{10} + \boldsymbol{1000}$ & $\boldsymbol{10} + \boldsymbol{2000}$ & $\boldsymbol{10} + \boldsymbol{5000}$ & $\boldsymbol{20} + \boldsymbol{5000}$ & $\boldsymbol{50} + \boldsymbol{5000}$ \\
                \midrule
                \specialrule{0em}{0.8pt}{0pt}
                \midrule
                Swapped Prediction
                 & $86.07$
                 & $86.52$
                 & $87.11$
                 & $91.44$
                 & $92.59$
                \\
                \midrule
                Cross-Entropy
                 & $85.63$
                 & $86.20$
                 & $86.72$
                 & $91.17$
                 & $92.38$
                \\
                KL Divergence
                 & $78.36$
                 & $79.62$
                 & $76.50$
                 & $87.56$
                 & $90.25$
                \\
                MSE
                 & $77.92$
                 & $78.20$
                 & $76.48$
                 & $86.30$
                 & $88.37$
                \\
                Cross-Entropy with Pseudo-Labeling
                 & $72.03$
                 & $72.80$
                 & $73.89$
                 & $85.65$
                 & $88.79$
                \\
                \bottomrule
            \end{tabular}
            \begin{tablenotes}
                \footnotesize
                \item The overall recognition accuracy with full supervision is $92.33\%$, achieved with $5000$ examples per device available for training. The composite data augmentation here used is $\left\{\operatorname{rot}_{0}, \, \operatorname{rot}_{\frac{1}{2}\pi}, \, \operatorname{rot}_{\pi}, \, \operatorname{rot}_{\frac{3}{2}\pi}\right\}$, followed by a $2$-segmented stochastic permutation operation. The loss scaling for Swapped Prediction takes $\alpha = 0$.
            \end{tablenotes}
        \end{threeparttable}}
\end{table*}

\subsection{Comparison for Different Consistency Forms}
Table \ref{Table: Comparison of Different Consistency Forms} gives a performance comparison of various consistency forms under different data conditions. It can be seen that Swapped Prediction is superior to other consistency forms, especially when labeled data is very limited, e.g., $\mathrm{M} = 10$. By carefully tuning $\lambda$, we can see that MSE achieves comparable performance with KL divergence, and both MSE and KL divergence likely deteriorate with more unlabeled samples provided. Meanwhile, we can also see that KL divergence is slightly worse than cross-entropy but significantly better than cross-entropy with pseudo-labeling. The above results are consistent with our previous analysis.

Even using only Swapped Prediction without considering other add-ons, such as EMA, we have achieved equivalent performance to full supervision using only $50$ labeled samples per class, with an overall recognition accuracy of more than $92\%$, while this is originally obtained by training with $5000$ examples per class. The result could be even better by cooperating with EMA or a stochastic permutation operation of more segments. The demand for labeled data is greatly reduced.

\subsection{Ablation Study of Exponential Moving Average}
Table \ref{Table: Ablation Study of Exponential Moving Average} compares SS-CSR's performance at different $\gamma$ settings. Given that Swapped Prediction is very effective, we can see that here EMA does not improve significantly for SS-CSR, although such effects often vary from dataset to dataset. In contrast to a slight performance improvement, we may be more concerned about how EMA improves training stability. Seen in Table \ref{Table: Ablation Study of Exponential Moving Average} again, we have listed a group of training statistics like $m/n$, where $m$ denotes a count for good results in every 10 consecutive trials, while $n$ is a count for poor results due to bad minima. The training stability is better with $m$ bigger and $n$ smaller. It can be seen that EMA with $\lambda = 0.9$ greatly improves training stability.

\begin{table}[htb]
    \renewcommand\arraystretch{1.15}
    \centering
    \caption{Ablation Study of Exponential Moving Average}
    \label{Table: Ablation Study of Exponential Moving Average}
    \begin{threeparttable}
        \begin{tabular}{cccc}
            \toprule
            \textbf{Decay}
             & $\boldsymbol{10} + \boldsymbol{1000}$ & $\boldsymbol{10} + \boldsymbol{2000}$ & $\boldsymbol{10} + \boldsymbol{5000}$ \\
            \midrule
            \specialrule{0em}{0.8pt}{0pt}
            \midrule
            w/o
             & $86.07 \ {\scriptscriptstyle 4/0}$
             & $86.52 \ {\scriptscriptstyle 3/0}$
             & $87.11 \ {\scriptscriptstyle 6/2}$
            \\
            \midrule
            $0.90$
             & $86.30 \ {\scriptscriptstyle 8/0}$
             & $86.72 \ {\scriptscriptstyle 7/0}$
             & $87.25 \ {\scriptscriptstyle 9/1}$
            \\
            $0.99$
             & $86.08 \ {\scriptscriptstyle 2/3}$
             & $86.59 \ {\scriptscriptstyle 3/5}$
             & $86.61 \ {\scriptscriptstyle 8/2}$
            \\
            \bottomrule
        \end{tabular}
    \end{threeparttable}
\end{table}

\subsection{Comparison with Other Methods}
To comprehensively prove effectiveness, we further validate SS-CSR on two real-world signal datasets, i.e., RadioML 2018.01A \cite{o2018over} and WIDEFT \cite{siddik2021wideft}, and compare it with recent semi-supervised algorithms of communication signal recognition, including SSRCNN \cite{dong2021ssrcnn}, MAT \cite{fu2022novel}, SimCLR \cite{liu2021self}, DCR \cite{fu2023semi}, and our past work, i.e., CDA \cite{wang2023semi}. Although some of them were designed with a single task in mind, only modulation recognition or RF fingerprinting, we still verify all. The same backbone network is used for all mentioned methods. Note that we have almost faithfully implemented every method and adjusted their associated hyper-parameters to be relatively optimal. The results are given in Table \ref{Table: Comparison with Other Semi-Supervised Methods for Modulation Recognition} for modulation recognition on RadioML 2018.01A and Table \ref{Table: Comparison with Other Semi-Supervised Methods for RF Fingerprinting} for RF fingerprinting on WIDEFT.

It can be seen that SS-CSR is superior to all other methods. In contrast to CDA, it is evident that SS-CSR wins for its better implementation of consistency-based regularization. Specifically, CDA suffers from pseudo-labeling that cannot take advantage of all unlabeled samples and simultaneously could be misled by those wrongly predicted samples but with high confidence. And our newly proposed Swapped Prediction has well solved these issues. As for DCR, similar to CDA, both have some drawbacks brought by pseudo-labeling, but DCR is more affected by its data augmentation. Specifically, DCR involves a cutout operation, randomly setting a section of sampling points in an input signal sample to zero. This operation significantly changes sample distributions and loses much symbol information. The performance of DCR is therefore affected. Meanwhile, we can also see that SSRCNN and MAT achieve relatively weak performance since they do not use any strong data augmentation and only consider noise injection. In addition, SimCLR achieves a promising result, but still far from ours. The experimental results above indicate that our proposed method has reached SOTA performance for SSL of communication signal recognition.

\begin{table}[htb]
    \renewcommand\arraystretch{1.15}
    \centering
    \caption{Comparison for Modulation Recognition on RadioML 2018.01A}
    \label{Table: Comparison with Other Semi-Supervised Methods for Modulation Recognition}
    \begin{threeparttable}
        \begin{tabular}{cccc}
            \toprule
            \multirow{2}{*}[-1.0ex]{\textbf{Method}}
             & \multicolumn{3}{c}{\textbf{Modulation Recognition}}
            \\
            \cmidrule(lr){2-4}
             & $\boldsymbol{10} + \boldsymbol{1000}$
             & $\boldsymbol{20} + \boldsymbol{1000}$
             & $\boldsymbol{50} + \boldsymbol{1000}$
            \\
            \midrule
            \specialrule{0em}{0.8pt}{0pt}
            \midrule
            SSCSR
             & $77.13$
             & $82.33$
             & $85.90$
            \\
            \midrule
            CDA
             & $76.05$
             & $79.22$
             & $82.36$
            \\
            DCR
             & $50.73$
             & $56.94$
             & $65.70$
            \\
            MAT
             & $43.66$
             & $52.89$
             & $62.05$
            \\
            SSRCNN
             & $36.95$
             & $39.52$
             & $49.80$
            \\
            SimCLR
             & $56.88$
             & $64.53$
             & $74.80$
            \\
            \bottomrule
        \end{tabular}
        \begin{tablenotes}
            \footnotesize
            \item Note that SS-CSR on RadioML 2018.01A takes $\alpha = 3.0$ and $\gamma = 0.90$. The composite data augmentation here used is $\left\{\operatorname{rot}_{0}, \, \operatorname{rot}_{\frac{1}{2}\pi}, \, \operatorname{rot}_{\pi}, \, \operatorname{rot}_{\frac{3}{2}\pi}, \, \operatorname{flip}_{\mathrm{h}}, \, \operatorname{flip}_{\mathrm{v}}\right\}$, followed by a $64$-segmented stochastic permutation operation.
        \end{tablenotes}
    \end{threeparttable}
\end{table}

\begin{table}[htb]
    \renewcommand\arraystretch{1.15}
    \centering
    \caption{Comparison for RF Fingerprinting on WIDEFT}
    \label{Table: Comparison with Other Semi-Supervised Methods for RF Fingerprinting}
    \begin{threeparttable}
        \begin{tabular}{cccc}
            \toprule
            \multirow{2}{*}[-1.0ex]{\textbf{Method}}
             & \multicolumn{3}{c}{\textbf{RF Fingerprinting}}
            \\
            \cmidrule(lr){2-4}
             & $\boldsymbol{10} + \boldsymbol{1000}$
             & $\boldsymbol{20} + \boldsymbol{1000}$
             & $\boldsymbol{50} + \boldsymbol{1000}$
            \\
            \midrule
            \specialrule{0em}{0.8pt}{0pt}
            \midrule
            SSCSR
             & $67.03$
             & $82.38$
             & $88.93$
            \\
            \midrule
            CDA
             & $65.08$
             & $81.03$
             & $87.25$
            \\
            DCR
             & $26.41$
             & $28.36$
             & $33.82$
            \\
            MAT
             & $19.63$
             & $30.53$
             & $53.07$
            \\
            SSRCNN
             & $10.07$
             & $13.85$
             & $43.66$
            \\
            SimCLR
             & $31.02$
             & $51.72$
             & $77.86$
            \\
            \bottomrule
        \end{tabular}
        \begin{tablenotes}
            \footnotesize
            \item Note that SS-CSR on WIDEFT takes $\alpha = 2.0$ and $\gamma = 0.00$ on WIDEFT. The composite data augmentation here used is $\left\{\operatorname{rot}_{0}, \, \operatorname{rot}_{\frac{1}{2}\pi}, \, \operatorname{rot}_{\pi}, \, \operatorname{rot}_{\frac{3}{2}\pi}, \, \operatorname{flip}_{\mathrm{h}}, \, \operatorname{flip}_{\mathrm{v}}\right\}$, followed by a $2$-segmented stochastic permutation operation.
        \end{tablenotes}
    \end{threeparttable}
\end{table}

% \begin{table}[htb]
%     \renewcommand\arraystretch{1.15}
%     \centering
%     \caption{Comparsion on Real-World Dataset, RadioML 2018.01A}
%     \label{Table: Comparsion on Modulation Recognition}
%     \resizebox{0.4\textwidth}{!}{
%         \begin{threeparttable}
%             \begin{tabular}{cccc}
%                 \toprule
%                 \textbf{Method}
%                  & $\boldsymbol{10} + \boldsymbol{1000}$ & $\boldsymbol{20} + \boldsymbol{2000}$ & $\boldsymbol{50} + \boldsymbol{5000}$ \\
%                 \midrule
%                 \specialrule{0em}{0.8pt}{0pt}
%                 \midrule
%                 SS-CSR
%                  & $86.05\%$
%                  & $86.05\%$
%                  & $86.05\%$
%                 \\
%                 \midrule
%                 Ours Previously
%                  & $86.05\%$
%                  & $86.05\%$
%                  & $86.05\%$
%                 \\
%                 DCR (Fu et al.)
%                  & $86.05\%$
%                  & $86.05\%$
%                  & $86.05\%$
%                 \\
%                 SSRCNN (Dong et al.)
%                  & $86.05\%$
%                  & $86.05\%$
%                  & $86.05\%$
%                 \\
%                 \bottomrule
%             \end{tabular}
%         \end{threeparttable}}
% \end{table}

\section{Conclusion} \label{Section: Conclusion}
This paper has investigated deep semi-supervised learning for communication signal recognition. We analyze different implementations of consistency-based regularization with their strengths and weaknesses and then propose a novel one, i.e., Swapped Prediction, which can effectively avoid such negative impacts from those wrongly predicted but high-confidence samples. Additionally, we emphasize it is necessary to use strong data augmentation in deep SSL and introduce EMA to improve performance and training stability further. The experimental results indicate that our proposed method significantly outperforms other competing ones. In future work, we expect to reduce further such dependency on labeled data for communication signal recognition and even achieve complete unsupervised learning.

\bibliographystyle{IEEEtran}
\bibliography{references}

\end{document}